\documentclass[a4paper,11pt]{article}
\usepackage{pos}
\usepackage{amsmath}
\usepackage{dsfont}
\usepackage{bm}

\title{Criteria of Renormalizability in Effective Field Theories}
\author*[a]{A.~M.~Gasparyan}
\author[a]{E.~Epelbaum}

\affiliation[a]{Fakult\"at f\"ur Physik und
	Astronomie, Institut f\"ur Theoretische Physik~II, Ruhr-Universit\"at Bochum,\\
   Bochum, D-44780, Germany}

\emailAdd{ashot.gasparyan@rub.de}
\emailAdd{evgeny.epelbaum@rub.de}

\abstract{Any effective field theory relies on power counting rules
	that allow one to perform a systematic expansion of calculated 
	quantities in terms of some soft scales. However, a naive 
	power counting can be violated due to the presence of
	various hard scales in a given scheme.
	A typical example of such a scale is an ultraviolet regulator.
	This issue is particularly challenging when the interaction is
	nonperturbative. The power counting is expected
	to be restored in the course of renormalization,
	that is by redefining bare low-energy constants
	in the effective Lagrangian.
	Whether this procedure eventually leads to a self-consistent framework is not a priory obvious.
	We discuss various criteria of renormalizability
	in application to nuclear chiral effective field theory and
	provide several instructive counterexamples.}

\FullConference{The 11th International Workshop on Chiral Dynamics (CD2024)\\
 26-30 August 2024\\
Ruhr University Bochum, Germany\\}

\begin{document}
\maketitle

\section{Introduction}
Nowadays,  the effective field theory (EFT) approach
is a standard and powerful tool in various areas of hadron physics.
It has also become popular in studies of two- and few-nucleon interactions,
in particular, due to success of chiral EFT with explicit pion degrees of freedom.
The original idea to implement chiral EFT in the few-nucleon sector 
was suggested by Weinberg in Refs.~\cite{Weinberg:1990rz,Weinberg:1991um}.
Since then, an enormous progress has been achieved in describing 
various empirical data and understanding theoretical aspects,
see Refs.~\cite{Epelbaum:2008ga,Machleidt:2011zz,Epelbaum:2019kcf,Hammer:2019poc}
for the reviews.

Chiral EFT is based on the corresponding power counting,
which implies an expansion of observables in the small quantity $Q=\frac{q}{\Lambda_b}$,
where the soft scale $q$ is determined by the pion mass $M_\pi$ and 
external 3-momenta $|\vec p|$, while the breakdown (hard) scale
is of the order of $\Lambda_b=500-700$~MeV.
In turn, the effective Lagrangian is expanded in powers of the quark mass $m_Q\sim M_\pi^2$
and the number of derivatives of the pion and nucleon fields.

To be able to make reliable predictions within an EFT,
one must make sure that the theory is renormalizable.
In the context of EFTs, this means that bare parameters of the effective
Lagrangian (possibly infinite) $C_i$ are expressed in terms of the finite renormalized
quantities 
\begin{align}
C_i=C_i^r+\delta C_i,
\label{Eq:C_i}
\end{align}
where $\delta C_i$ are counter terms.
The EFT expansion for observables is then formulated in terms of
the renormalized quantities.
In the case of the few-nucleon dynamics, the renormalization procedure is complicated by the
nonperturbative nature of the interaction,
which requires resummation of the leading-order potential contributions~\cite{Weinberg:1990rz,Weinberg:1991um}.
This is the reason why the renormalizability of chiral EFT in the two- and few-nucleon sectors is
not yet fully understood, although a significant progress has been made,
see Refs.~\cite{Hammer:2019poc,Epelbaum:2018zli,Gasparyan:2021edy,Gasparyan:2023rtj} for recent discussions.

In this talk, we discuss renormalizability criteria of an effective field theory in applications to 
chiral EFT for the two-nucleon system and consider some toy-model examples illustrating
those constraints.

\section{Regularization and renormalization}
The nonperturbative treatment of the leading-order (LO)
chiral nucleon-nucleon ($NN$) potential including the one-pion-exchange
contribution requires introducing a regulator for an infinite set of divergent
loop diagrams, typically, in the form of a cutoff $\Lambda$.
Renormalization of the amplitude in the spirit of a perturbative quantum field theory
would then require an infinite number of counter terms.
This would allow one to remove the regulator by setting $\Lambda\to\infty$
and introduce instead a renormalization scale(s) $\mu$.
In practice, incorporating an infinite number of contact interactions explicitly
is not feasible. Alternatively, one can keep the cutoff $\Lambda$ finite
and treat it as the renormalization scale (or the quantity related to it)
in the sense that it fixes implicitly the values of higher-order renormalized
coupling constants. 

One can explicitly implement the idea of interpreting $\Lambda$ as the renormalization
scale by formally splitting the effective Lagrangian into the regularized part and remaining terms:
\begin{align}
	\mathcal{L}(x)=\mathcal{L}_\Lambda(x)+\delta_\Lambda\mathcal{L}(x),
\end{align}
and expanding the observables in powers of $\delta_\Lambda\mathcal{L}(x)$~\cite{Gasparyan:2021edy, Gasparyan:2023rtj}.
The $\Lambda$-independence of the full Lagrangian, $\frac{d\mathcal{L}}{d\Lambda}=0$,
implies the formal (before renormalization) cutoff independence of the scattering amplitude
($T$-matrix) $\frac{dT}{d\Lambda}=0$.
This is a manifestation of the renormalization scale invariance of the EFT.
Obviously, $\Lambda$ independence at each EFT order is fulfilled only approximately
up to higher orders in $1/\Lambda$.
The fact that taking into account perturbative corrections in $\delta_\Lambda\mathcal{L}(x)$
significantly reduces the cutoff dependence of the amplitude
also after renormalization was confirmed by calculations in Refs.~\cite{Gasparyan:2021edy, Gasparyan:2023rtj}.

For locally regularized long-range potentials, such as the one-pion-exchange potential,
one can expand $\delta_\Lambda\mathcal{L}(x)$ in powers of $1/\Lambda$
obtaining the series of local interactions and restoring
the usual form for the effective Lagrangian.

Note that convergence of the perturbative expansion of a field theory
depends on the renormalization scale also in more fundamental quantum field theories
such as quantum electrodynamics (at some scale, there appears the Landau pole \cite{Landau:1954cxv})
and quantum chromodynamics (the problem of the renormalization scale ambiguity~\cite{Stevenson:1980du,Brodsky:1982gc}).

As already mentioned, the cutoff value cannot be chosen
to be very large $\Lambda\gg\Lambda_b$ (see, however, Sec.~\ref{Sec:RGI}). On the other hand,
one cannot set $\Lambda$ to be of the order of the soft scale
(which is possible, e.g., in the scheme with purely perturbative
pions, where one can also use  
dimensional regularization \cite{Kaplan:1998we}) as it would lead to 
appearance of cutoff artifacts.
Therefore, the natural option is to choose the cutoff of the order of the hard scale
$\Lambda\sim\Lambda_b$.
This can be understood qualitatively \cite{Lepage:1997cs,Gegelia:1998iu,Gegelia:2004pz,Epelbaum:2006pt}
and quantitatively \cite{Gasparyan:2021edy} by observing
that in the iterations of the LO interactions, positive powers of
$\Lambda$ in the amplitude
are compensated by negative powers of the hard scale originating 
from the potential.
Thus, all terms in the series are of the same EFT order $O(Q^0)$.
Nevertheless, after the resummation, the amplitude can be enhanced due to 
nonperturbative effects.

For the next-to-leading-order (NLO) potential and beyond, positive powers of $\Lambda$ from loop integrals
appear in places where one expects soft scales from dimensional arguments,
thereby violating power counting. These terms originate from the regions
of momenta $p\sim\Lambda$.
In this case, one can expect that after renormalization, the lower-order counter terms,
as defined in Eq.~\eqref{Eq:C_i}, will cancel the power-counting breaking contributions, 
since there is only a finite number of the uncompensated positive powers of $\Lambda$. 
This turns out to be a rather complicated procedure.
In practice, one typically performs implicit renormalization, which means
that all bare coupling constants are fitted for each EFT order separately.
One concludes then that the power counting is fulfilled by looking
at the convergence pattern of the EFT expansion.
It is, however, difficult to identify the individual contributions
from various EFT orders in such a scheme, which makes applications to
different processes less transparent and complicates maintaining 
the symmetries of the underlying theory after renormalization.

Nevertheless, the implicit-renormalization approach can be trusted as long as
one can show that it is equivalent to an explicit scheme, i.e.,
it can be proven that explicit renormalization in the sense of Eq.~\eqref{Eq:C_i},
can in principle be realized.

Recently, the renormalizability of chiral EFT in the $NN$ sector has been
rigorously proven at NLO for a rather general set of ultraviolet regulators
provided certain requirements on the short-range part of the LO potential are met \cite{Gasparyan:2021edy, Gasparyan:2023rtj}.
Upon analyzing the proof, it becomes clear that it is the unique feature of the field theoretical 
approach that makes nuclear chiral EFT renormalizable.
This allows us to formulate the criteria of renormalizability and
provide several instructive counterexamples.

\section{Renormalization of the nucleon-nucleon amplitude at NLO. }
\label{Sec:Renormalization_NLO}
In this section we briefly discuss how renormalization can be
performed explicitly 
in the $NN$ sector within nuclear chiral EFT at next-to-leading order.
The LO off-shell $NN$ amplitude satisfies the partial-wave Lippmann-Schwinger equation
$T_0=V_0+V_0GT_0$, or explicitly:
\begin{align}
	T_{0}(p',p;p_\text{on})&=V_0(p',p)+
	\int \frac{p''^2 dp''}{(2\pi)^3}
	V_0(p',p'')
	G(p'';p_\text{on})
	T_{0}(p'',p;p_\text{on}),\nonumber\\
	G(p''; p_\text{on})&=\frac{m_N}{p_\text{on}^2-p''^2+i \epsilon},
	\label{Eq:LS_equation}
\end{align}
where the matrix form is assumed for coupled channels.
The interaction is characterized by the LO and NLO potentials (of order $O(Q^0)$ and $O(Q^2)$) $V_0$ and $V_2$
that consist of the long-range $V_{0,\text{L}}$, $V_{2,\text{L}}$ and short-range $V_{0,\text{S}}$, $V_{2,\text{S}}$ parts:
\begin{align}
	V_0(p',p)=V_{0,\text{L}}(p',p)+V_{0,\text{S}}(p',p),\qquad
	V_2(p',p)=V_{2,\text{L}}(p',p)+V_{2,\text{S}}(p',p),
\end{align}
and (at least) the LO potential is regulated by a cutoff $\Lambda$.

The explicit solutions for the LO and (perturbative) NLO amplitudes 
can be represented as
\begin{align}
	&T_0=(\mathds{1}-V_0 G)^{-1}\, V_0,\label{Eq:T0_NP}\\
	&T_2=(\mathds{1}-V_0 G)^{-1} \, V_2 \,(\mathds{1}-G V_0)^{-1}.\label{Eq:T2_NP}
\end{align}
One can also expand these expressions in $V_0$ if the series are convergent
\begin{align}
	&T_0=\sum_{n=0}^{\infty}T_0^{[n]},\qquad  T_0^{[n]}=V_0 (G V_0)^n , \label{Eq:T0}\\
	&T_2=\sum_{m,n=0}^{\infty}T_2^{[m,n]},\qquad  T_2^{[m,n]}=(V_0 G)^m V_2 (G V_0)^n.\label{Eq:T2}
\end{align}
We treat first the perturbative case when the above series converge
and then discuss further complications due to nonperturbative effects.

\subsection{Perturbative treatment}
\label{Sec:perturbative}
Consider the one-loop contribution to $T_2$ for on-shell external momenta:
$T_2^{[0,1]}= V_2 G V_0$ or
\begin{align}
T_2^{[0,1]}(p_\text{on})\coloneqq T_2^{[0,1]}(p_\text{on},p_\text{on};p_\text{on})= \int \frac{p''^2 dp''}{(2\pi)^3}
 V_2(p_\text{on},p'') G(p'';p_\text{on}) V_{0}(p'',p_\text{on}).
\end{align}
At large momenta $p''$, the potentials behave as (here and in what follows, we neglect logarithmic corrections)
\begin{align}
	V_{0}(p'',p_\text{on})\sim \frac{1}{\Lambda_b^2},\qquad V_{2}(p_\text{on},p'')\sim \frac{p''^2}{\Lambda_b^4},
\end{align}
so that 
\begin{align}
T_2^{[0,1]}(p_\text{on})\sim \frac{m_N \Lambda^3}{\Lambda_b^6}\ne O(Q^2),
\end{align}
which violates the power counting.
To remedy this, one can perform a subtraction of the amplitude, e.g., at $p_\text{on}=0$.
The estimation of subtracted integrals is based on the following bounds
on the subtracted potentials for large momenta $p'>p$ obtained in Ref.~\cite{Gasparyan:2021edy}:
\begin{align}
\Delta_p^{(n)} V_\alpha(p',p)\coloneqq
V_\alpha(p',p)-\sum_{i=0}^{n}\frac{\partial^i V_\alpha(p',p)}{i!(\partial p)^i}\bigg|_{p=0}p^i
\sim \left(\frac{p}{p'}\right)^{n+1}V_\alpha(p',p),\qquad \alpha=0,2.\label{Eq:subtraction}
\end{align}
This behavior of the subtracted potentials is a direct consequence of the
structure of interactions derived within chiral EFT.
Particularly, the long-range part of the potential is local as it originates from
the one- and multiple-pion exchange contributions.
To be more precise, the positions of the pion-exchange singularities of the plain-wave potential
are functions of the momentum transfer squared $\vec q\,^2=(\vec p\,'-\vec p)^2$
and not of $\vec p$ and $\vec p\,'$ separately (a very general form of such a potential
satisfying Eq.~\eqref{Eq:subtraction} is considered in Ref.~\cite{Gasparyan:2021edy}).

Bounds in Eq.~\eqref{Eq:subtraction} lead to suppression of large loop momenta.
Therefore, the subtracted renormalized amplitude fulfills the expected power counting
if we choose $\Lambda\sim\Lambda_b$ (or rather $m_N\Lambda \sim\Lambda_b^2$):
\begin{align}
\mathds{R}(T_2^{[0,1]})(p_\text{on})=T_2^{[0,1]}(p_\text{on})-T_2^{[0,1]}(0)
\sim \frac{m_N \Lambda}{\Lambda_b^6}p_\text{on}^2
\sim \frac{p_\text{on}^2}{\Lambda_b^4}\sim O(Q^2).
\end{align}
This subtraction is equivalent to adding a constant counter term to $V_0$,
i.e. redefining the renormalized LO contact interactions.
Now, if we make one more iteration of $V_0$ and consider
$\mathds{R} (T_2^{[0,1]}) G V_0$, it is obvious
that the resulting integral will still violate the power counting
for the same reasons as the one-loop term.
Making another subtraction,
we can again restore the power counting by adding another counter term
of the same form to $V_0$.
Proceeding this way recursively, one can renormalize all 
terms $T_2^{[0,n]}=V_2 (G V_0)^n$ with just one type of the counter term.
Unfortunately, the situation is more complicated for the loop contributions
of the form $T_2^{[m,n]}=(V_0 G)^m V_2 (G V_0)^n$ with $m\ne 0$, $n\ne 0$,
because one can obtain such a diagram by adding one more loop with $V_0$
either from $T_2^{[m-1,n]}$ or from $T_2^{[m,n-1]}$.
This issue is analogous to the problem of overlapping divergencies
in quantum field theory.
It can be handled by applying the Bogoliubov-Parasiuk-Hepp-Zimmermann (BPHZ) subtraction scheme
\cite{Bogoliubov:1957gp,Hepp:1966eg,Zimmermann:1969jj} and partitioning
the multidimensional integration region into appropriate sectors.
It turns out that in this case the renormalization can be carried out as well,
and the renormalized amplitude satisfies the intended power counting
$\mathds{R}(T_2^{[m,n]})=O(Q^2)$ \cite{Gasparyan:2021edy}.
The required counter terms again correspond to the momentum-independent
contact interactions
already present in the LO potential.

\subsection{Nonperturbative effects}
\label{Sec:nonperturbative}
In the $NN$ sector, the dynamics in the partial waves $^1S_0$, $^3S_1-{^3D_1}$ and $^3P_0$
is essentially nonperturbative, which means
that the series in Eqs.~\eqref{Eq:T0},~\eqref{Eq:T2} do not converge or converge extremely slowly.
In this case, two more steps must be done.

First, it is possible to resum the series for the renormalized ampitude $\mathds{R}(T_2)$
in a closed form~\cite{Gasparyan:2023rtj}:
\begin{align}
	\mathds{R}( T_2)(p_\text{on}) = \sum_{m,n}\mathds{R}( T_2^{[m,n]})(p_\text{on})
	=  T_2(p_\text{on})
	+\delta C\psi(p_\text{on})^2,
	\label{Eq:nonperturbative_subtraction_ij}
\end{align}
where the vertex function $\psi$ is defined as 
\begin{align}
	\psi(p_\text{on}) &= 1+\int \frac{p^2 dp}{(2\pi)^3} G(p;p_\text{on}) T_{0}(p, p_\text{on};p_\text{on}).
	\label{Eq:psi_explicit}
\end{align}
The counter term constant (subtraction is needed only for the $S$-waves) is given by
\begin{align}
	\delta C=
	-\frac{T_2(0)}{\psi(0)^2},
	\label{Eq:delta_C_a}
\end{align}
so that $\mathds{R}(T_2)$ satisfies the renormalization condition
\begin{align}
	\mathds{R}(T_{2})(0)=0.
	\label{Eq:renormalization_condition_Swave}
\end{align}

The second step is to employ the Fredholm method for solving integral equations
and represent the amplitudes $T_0$, $T_2$ and the vertex function $\psi$
as the ratio of some quantities for which the perturbative expansion in $V_0$
is convergent and certain powers of the Fredholm determinant $D_0$~\cite{Gasparyan:2023rtj}.
The Fredholm determinant is a function of the on-shell
momentum $D=D(p_\text{on})$ and contains all the nonperturbative dynamics
including the enhancement of the amplitude due to the presence of a (quasi-) bound state,
as is the case in the channels $^1S_0$ and $^3S_1-{^3D_1}$.
This approach allows one to apply all the arguments about renormalizability of 
chiral EFT also in the nonperturbative regime.

The only additional constraint arises due to the inverse of $\psi(0)^2$ in 
Eq.~\eqref{Eq:delta_C_a}. If $\psi(0)$ is too small, the counter term $\delta C$
becomes unnaturally large.
Then the amplitude $\mathds{R}(T_2)$ explodes away from the threshold,
and renormalizability breaks down.
An explicit calculation reveals that as long as the cutoff does not exceed the hard
scale $\Lambda\lesssim\Lambda_b$, such a situation never occurs and the counter terms
have natural size.

To summarize, 
one can formulate three remormalizability criteria for nuclear chiral EFT (and analogous theories):
\begin{itemize}
	\item Locality of the long-range forces.
	\item Cutoff of the order of the hard scale $\Lambda\sim\Lambda_b$.
	\item Natural size of (finite parts of) the counter
          terms. Note that we are talking here about the (finite) counter terms
	that absorb the uncompensated positive powers of $\Lambda$ for $\Lambda\sim\Lambda_b$.
\end{itemize}
In the subsequent sections, we provide several counterexamples, where
these criteria are violated to illustrate the fact that the renormalizability
is not a trivial property of a theory and  is not always guaranteed.

\section{Nonlocal separable interactions}
\label{Sec:nonlocal}
In this section we discuss the toy model where the locality condition for the long-range forces
is not fulfilled. We consider a separable potential form for both LO and NLO interactions.
Separable models are frequently employed in the literature for illustrative purposes, 
see Refs.~\cite{Epelbaum:2017byx,Peng:2021pvo} for recent usage.

The issue of remormalization of various modifications of separable interactions
is analyzed in detail in Ref.~\cite{Jacobi:2025aaa}. Here, we focus on one specific model
to demonstrate the situation when renormalization fails.
In this model, both $V_0$ and $V_2$ have a simple separable form.
The LO potential is purely short-range and the NLO potential
contains the long-range part that resembles the two-pion exchange
contribution in the nucleon-nucleon chiral EFT:
\begin{align}
	V_0(p',p)&=C_0 F_\Lambda(p')F_\Lambda(p),\\
	V_2(p',p)&=C_2 (p'^2+p^2)\frac{p'^2 p^2}{(M_\pi^2+p'^2)(M_\pi^2+p^2)}
	F_\Lambda(p')F_\Lambda(p),
\end{align}
where $F_\Lambda$ is some regulator with the cutoff $\Lambda\sim\Lambda_b$,
and the coupling constants are of natural size $C_0\sim\Lambda_b^{-2}$, $C_2\sim\Lambda_b^{-4}$.
Of course, $V_0$ can contain also the long-range part corresponding to the one-pion exchange interaction,
but this is irrelevant to further analysis.

To understand where the renormalization problem arises,
consider a one-loop contribution to the on-shell NLO amplitude
and estimate it as in Sec.~\ref{Sec:perturbative}:
\begin{align}
	V_0 G V_2&\sim \frac{1}{\Lambda_b^2}
\frac{p_\text{on}^2}{(M_\pi^2+p_\text{on}^2)}
	\sim O(Q^0).
\end{align}
This term violates the power counting and requires a
counter term that is nonlocal (long-range). 
Moreover, it has a long-range structure which, as we assumed, 
is not present in the LO potential, as it corresponds to
the "two-pion-exchange" contribution.
Similarly, we would have to introduce such nonlocal counter
terms for all higher-order interactions from the very beginning,
which is impossible within an EFT approach, where
one performs an expansion order by order.

This failure of renormalizability is a direct consequence
of the violation of bounds~\eqref{Eq:subtraction}.
In gerneral, they do not hold for non-local long-range interactions.

\section{Exceptional cutoffs in the "renormalization-group invariant" scheme}
\label{Sec:RGI}
Next, we consider the situation when nonperturbative effects hinder
renormalizability (the values of the LECs become unnatural). 
A typical example is the $^3P_0$ channel in the
$NN$ system within the framework of chiral EFT. 
We assume that the LO potential consists of the one-pion-exchange
contribution and a contact term regularized, e.g., by a nonlocal
or sharp regulator with a cutoff $\Lambda$. We want to renormalize
the NLO amplitude, which contains the two-pion-exchange potential
and higher order contact interactions, according to the prescription
provided in Sec.~\ref{Sec:nonperturbative}.
Then, it turns out that already for cutoffs slightly larger than the 
hard scale $\Lambda\gtrsim\Lambda_b$
(the exact value depends on the type of a regulator)~\cite{Gasparyan:2023rtj},
the counter term determined by Eq.~\eqref{Eq:delta_C_a}
becomes too large to ensure renormalizability of the amplitude.

There is an approach in the literature stating that
an EFT expansion must work equally well not only 
for $\Lambda\sim\Lambda_b$ but also for 
(much) larger values of $\Lambda$, while the number of
the low energy constants at each EFT order is kept finite
(at least, in a single partial wave), see Rev.~\cite{Hammer:2019poc} for a review.
In particular, the following approximate "renormalization-group (RG) invariance"
for the $T$-matrix at the EFT order $\mathcal{V}$ is postulated:
\begin{align}
	\frac{\Lambda}{T^{(\mathcal{V})}(Q,\Lambda)}
	\frac{{\mathrm{d}} T^{(\mathcal{V})}(Q,\Lambda)}{{\mathrm{d}}\Lambda}
	=  O\left(\frac{Q^{{\mathcal{V}}+1}}{M_{\text{hi}}^{\mathcal{V}}\Lambda}\right),\label{Eq:RGI}
\end{align}
where $Q$ is the typical momentum in the system
and $M_{\text{hi}}$ is the EFT expansion breakdown scale ($\Lambda_b$ in our notation). 
Equation~\eqref{Eq:RGI} implies the existence of the $\Lambda\to\infty$ limit.
Moreover, one requires that the amplitude be independent of the
functional form of the regulator for sufficiently large values of $\Lambda$~\cite{Song:2016ale}.
Here, we do not discuss critical arguments against such a scheme
from a conceptual point of view, see Refs.~\cite{Epelbaum:2009sd,Epelbaum:2018zli,Epelbaum:2020maf,Epelbaum:2021sns},
but rather focus on a particular aspect of renormalizability issues.
One of the motivations for Eq.~\eqref{Eq:RGI} is the quantum-mechanical case
of the singular attractive potentials~\cite{Frank:1971xx}, an example of which is
the unregulated LO one-pion-exchange potential in the $^3P_0$ channel, see~\cite{Nogga:2005hy}.
Beyond leading order, this analogy no longer holds as, e.g., the two-pion-exchange potential 
can become repulsive, and its nonperturbative treatment 
turns impossible~\cite{PavonValderrama:2005wv,PavonValderrama:2005uj,Zeoli:2012bi,vanKolck:2020llt}.

The "RG-invariance" in the form of Eq.~\eqref{Eq:RGI} is achieved by 
introducing more contact interactions as compared to naive dimensional analysis.
In the $^3P_0$ channel, there appear one contact term at LO and two contact terms at NLO~\cite{Long:2011qx}.
Nevertheless, for a certain "exceptional" value of the cutoff $\Lambda\gtrsim\Lambda_b$, the NLO constants
become extremely large (infinite for a specific $\Lambda$ value), and the scattering amplitude explodes.
The region around the "exceptional" cutoff where the renormalization breaks down
is extremely narrow, around $0.1$~MeV, which makes it difficult to notice
in numerical calculations. This carries a potential risk in practical applications.

Since in the "RG-invariant" scheme one is interested in the cutoffs $\Lambda>\Lambda_b$,
one could ignore this "exceptional" region and proceed with larger values of $\Lambda$.
Unfortunately, there is an infinite number of such "exceptional" cutoffs on the real axis
due to the singular nature of the unregulated one-pion-exchange potenatial~\cite{Gasparyan:2022isg}.
Therefore, there is no continuous flow of the solution of Eq.~\eqref{Eq:RGI}
towards $\Lambda=\infty$ (or from $\Lambda=\infty$ to
$\Lambda\sim\Lambda_b$), and no $\Lambda\to\infty$ limit exists.
Consequently, the naive requirement of the approximate (up to $1/\Lambda$ corrections)
independence of the $T$-matrix of the functional form of the regulator 
and the value of the cutoff must be modified.

Recently, two attempts have been made to remedy the above problem \cite{Peng:2024aiz,Yang:2024yqv}.
The approach of Ref.~\cite{Peng:2024aiz} suggests that one modifies
the renormalization conditions in the vicinity of the "exceptional" cutoffs.
This leads, however, to a discontinuous behavior of the scattering amplitude
as a function of the cutoff.
In Ref.~\cite{Yang:2024yqv}, one modifies the functional form of the regulator
close to "exceptional" cutoffs.
In turn, this modification of the regulator is discontinuous in $\Lambda$,
although this feature is not mentioned in the paper explicitly.
In fact, the following argument shows that one cannot avoid an "exceptional"
cutoff $\bar\Lambda$ by a continuous (as a function of $\Lambda$) change of either the renormalization conditions
$renorm(\Lambda)$ or the functional form of the regulator $F_\Lambda(\Lambda,p)$ in a finite region around such a cutoff 
$\Lambda\in(\bar\Lambda-\delta\Lambda,\bar\Lambda+\delta\Lambda)$.
As $\Lambda$ passes from $\bar\Lambda-\delta\Lambda$ to $\bar\Lambda+\delta\Lambda$,
the position of $\bar\Lambda\big(renorm(\Lambda),F_\Lambda(\Lambda,p)\big)$ makes a closed continuous path
around its starting value, and, therefore, at some point $\Lambda_0$ they intersect:
$\Lambda_0=\bar\Lambda\big(renorm(\Lambda_0),F_{\Lambda_0}(\Lambda_0,p)\big)$.
Discontinuities of the RG flow makes the usual EFT interpretation in terms of
Wilsonian RG rather unclear. 

Another unfortunate aspect of the prescriptions 
based on modifications of the renormalization conditions or
a functional form of the regulator
is that one is forced to design the LO interaction according
to information about possible (unknown at the beginning) issues appearing at higher orders.
As one goes beyond NLO or considers different processes,
new "exceptional" cutoffs will show up located at different positions,
which will require further modification of the treatment of the LO interaction.

For the case of "exceptional" cutoffs, two conditions from Sec.~\ref{Sec:Renormalization_NLO} are violated:
$\Lambda\sim\Lambda_b$ and naturalness of the LECs, which leads
to failure of renormalizability.

\section{Nonperturbative effects with short-range separable interactions}
In this section we demonstrate the appearance of nonperturbative effects
in renormalization using again a simple separable model.
We choose purely short-range (pionless) interactions and artificially make them
pathological to illustrate the key idea.
The LO potential is the same as in Sec.~\ref{Sec:nonlocal}
with the cutoff $\Lambda=\Lambda_b$:
\begin{align}
	V_0(p',p)&=C_0 F_{\Lambda_b}(p')F_{\Lambda_b}(p).
\end{align}
The regulator is taken in the following sharp form:
\begin{align}
	F_\Lambda(p)=\theta(\Lambda -p)\left(1-\frac{a}{\Lambda^2}p^2\right),
\end{align}
where $a$ is the parameter representing the short-range ambiguity of the 
potential (or a promotion of the higher-order interaction).
The NLO potential is simply
\begin{align}
	V_2(p',p)=C_2(p^2+p'^2).
\end{align}
The LO and NLO on-shell amplitudes, $T_0$ and $T_2$ are given by
\begin{align}
	T_0(p_\text{on})=\frac{C_0}{1-C_0\,\Sigma(p_\text{on})},\qquad
	T_2(p_\text{on})=2 C_2\psi(p_\text{on})\psi_2(p_\text{on}),
\end{align}
where
\begin{align}
\psi(p_\text{on})=1+\frac{C_0\Sigma_0(p_\text{on})}{1-C_0\,\Sigma(p_\text{on})},\qquad
\psi_2(p_\text{on})=p_\text{on}^2+\frac{C_0\Sigma_2(p_\text{on})}{1-C_0\,\Sigma(p_\text{on})},
\end{align}
and
\begin{align}
	\Sigma(p_\text{on})&=\int \frac{p^2 dp}{(2\pi)^3}F_{\Lambda_b}(p)^2	G(p;p_\text{on}), \nonumber\\
	\Sigma_0(p_\text{on})&=\int \frac{p^2 dp}{(2\pi)^3}F_{\Lambda_b}(p)	G(p;p_\text{on}), \nonumber\\
	\Sigma_2(p_\text{on})&=\int \frac{p^2 dp}{(2\pi)^3}F_{\Lambda_b}(p) p^2	G(p;p_\text{on})
	=\Sigma_2(0)+\Sigma_0(p_\text{on})p_\text{on}^2.
\end{align}
Again, $T_2$ violates the power counting due to nonzero $\Sigma_2(0)$.
To remove the power-counting breaking contribution, we add
to the NLO potential $V_2$ the counter term $\delta C_0$ given by Eq.~\eqref{Eq:delta_C_a}
and obtain the followning term in the NLO $T$-matrix:
\begin{align}
	\delta T_2(p_\text{on})=-2C_2\frac{\psi_2(0)}{\psi(0)}\psi(p_\text{on})^2.
\end{align}
In the case of separable interactions, there are no overlapping "divergencies",
and the restoration of the power counting is easy to verify:
\begin{align}
	\mathds{R}(T_2)(p_\text{on})=
	2 C_2 \frac{\psi(p_\text{on})}{\psi(0)}
\left[\psi_2(p_\text{on})\psi(0)-\psi_2(0)\psi(p_\text{on})\right].
\end{align}

Using the explicit expressions
\begin{align}
	\Sigma(0)=\frac{m_N\Lambda_b}{(2\pi)^3}\left(-1+\frac{2a}{3}-\frac{a^2}{5}\right),\qquad
	\Sigma_0(0)=\frac{m_N\Lambda_b}{(2\pi)^3}\left(-1+\frac{a}{3}\right),
\end{align}
we choose the critical values of $C_0$ and $a$ leading to 
the condition $\psi(0)=0$:
\begin{align}
	C_0=-\frac{(2\pi)^3}{m_N\Lambda_b},\qquad
	a=\frac{5+\sqrt{205}}{6}\approx 3.22.\label{Eq:C_0_a}
\end{align}
Note that in this case, $C_0 \Sigma_0(0)\approx 0.93$, i.e., there is no bound state, and
the expansion of $T_0$ and (urenormalized) $T_2$ in terms of $V_0$ is formally convergent,
although each individual term in the expansion of $T_2$ violates the power counting.
Interestingly, after the renormalization, $\mathds{R}(T_2)$, which obeys the power counting,
does not converge when expanded in $V_0$.
Moreover, $\mathds{R}(T_2)$ explodes outside a small energy region around the threshold
(zero-range region if the values of $C_0$ and $a$ are taken exactly as in Eq.~\eqref{Eq:C_0_a}),
and renormalization does not work.

To summarize, we have demonstrated how nonperturbative effects can
lead to unexpected results and failure of renormalizability.

\section{Conclusion}
We have analyzed recent results on renormalizability of 
chiral effective field theory, specifically in the nucleon-nucleon sector,
and formulated three conditions needed for an EFT to be renormalizable:
\begin{itemize}
	\item The long-range part of the chiral forces are local, which is fulfilled automatically if
	a theory is based on an effective Lagrangian with the pionic degrees of freedom.
	\item The ultraviolet regulator depends on a cutoff of the order of the hard scale $\Lambda\sim\Lambda_b$.
	\item The finite parts of the counter terms that absorb the uncompensated positive powers of $\Lambda$
	are of natural size in terms of $\Lambda_b$.
\end{itemize}

We have provided counterexamples demonstrating how violating one or several of the
above conditions can lead to failure of renormalizability.

First, we discussed the model with a nonlocal separable long-range interaction,
which is at odds with the first renormalizability condition.
In this model, one cannot absorb the power-counting breaking contributions
by local contact interactions. Moreover, even with nonlocal counter terms,
one has to include in the LO potential the ones corresponding to all higher-order interactions
from the very beginning, which is impossible within the paradigm of an EFT.

Second, we analyzed the appearance of "exceptional" cutoffs in the so-called
"renormalization-group invariant" nuclear chiral EFT when applied to the $^3P_0$ channel
of the $NN$ interaction. At such cutoffs, the scattering amplitude explodes,
and the renormalization breaks down due to the violation of the second
and third renormalizability conditions.

Finally, we constructed a separable model with a purely short-range interaction
and fine-tuned its parameters to create a pathological situation when the third
renormalizability condition is maximally violated similarly to the case of "exceptional" cutoffs.
As expected, renormalizability fails for this model too.

These examples illustrate the rigor and predictive power of EFTs in the few-nucleon
physics compared to more phenomenological approaches.

\acknowledgments
This work has been supported in part by the European Research Council (ERC AdG NuclearTheory, grant No. 885150), 
by the MKW NRW under the funding code NW21-024-A, by JST ERATO (Grant No. JPMJER2304) and by JSPS KAKENHI (Grant No. JP20H05636).

\bibliographystyle{JHEP}
\bibliography{5.8}

\end{document}